%% file: main.tex
\documentclass{article}
\usepackage{spconf,amsmath,graphicx}

\title{Spiking Structured State Space Model for Monaural Speech Enhancement}
\twoauthors {Yu Du\sthanks{Thanks to the National Natural Science Foundation of China (61836004, 62236009).}} {Tsinghua University\\Department of Precision Instrument\\Beijing} 
{Xu Liu\sthanks{Co first author\\Thanks to the STI 2030—Major Projects2021ZD0200300.}, Yansong Chua\sthanks{Corresponding author\\Thanks to the STI 2030—Major Projects2021ZD0200300.}} {China Nanhu Academy \\of Electronics and Information Technology\\Jiaxing, China.}

\usepackage{bibentry}
\usepackage{algpseudocode}
\usepackage{amsfonts}
\usepackage{xspace}
\usepackage{booktabs}

\usepackage{subfig}
\usepackage{subfloat}
\usepackage{graphicx}
\usepackage{amsmath}
\usepackage[capitalize]{cleveref}
\makeatletter
\DeclareRobustCommand\onedot{\futurelet\@let@token\@onedot}
\def\@onedot{\ifx\@let@token.\else.\null\fi\xspace}

\makeatother

\begin{document}
%
\maketitle
\input{sections/abstract}
\input{sections/keywords}
\input{sections/intro}
\input{sections/pre}

\input{sections/method}

\input{sections/exp}

\input{sections/con}

\small
\bibliographystyle{IEEEbib}
\bibliography{strings,refs}

\end{document}

%% file: sections/abstract.tex
\begin{abstract}

Speech enhancement seeks to extract clean speech from noisy signals. Traditional deep learning methods face two challenges: efficiently using information in long speech sequences and high computational costs. To address these, we introduce the Spiking Structured State Space Model (Spiking-S4). This approach merges the energy efficiency of Spiking Neural Networks (SNN) with the long-range sequence modeling capabilities of Structured State Space Models (S4), offering a compelling solution. Evaluation on the DNS Challenge and VoiceBank+Demand Datasets confirms that Spiking-S4 rivals existing Artificial Neural Network (ANN) methods but with fewer computational resources, as evidenced by reduced parameters and Floating Point Operations (FLOPs).
\end{abstract}

%% file: sections/keywords.tex

%% file: sections/intro.tex
\vspace{-2mm}
\section{Introduction}
\label{sec:intro}
Speech enhancement aims to separate clean speech signals from noisy backgrounds or communication embedded with noise. This task is notoriously difficult due to two main factors. First, both speech and noise signals are dynamic and change over time, and they are unrelated and independent from each other. Second, speech signals often carry meaningful information within extended sequences. Therefore, extracting valuable insights from these lengthy sequences has consistently presented a significant challenge.

The intricate interplay between speech and noise, coupled with varying environmental conditions, adds complexity. Striking a balance between noise reduction and preserving speech clarity is a challenge for algorithm design. Various techniques, ranging from traditional filters to modern machine learning, have been employed to address this problem. Deep neural networks, including Feed Forward Networks~(FFNs), Recurrent Neural Networks~(RNNs), Convolutional Neural Networks~(CNNs), and Transformers, have been extensively explored in this domain~\cite{ochieng2022deep}. Among these, Unet-like structures have achieved state-of-the-art performance~\cite{zhao2022frcrn, zhao2021monaural}.

In addition, RNN-based solutions, face challenges in modeling long-term dependencies~\cite{erdogan2015phase, weninger2014discriminatively} and transformers are limited by memory and computation efficiency constraints ~\cite{subakan2021attention, zhao2020monaural}. Recently, the structured state space model (SSM) series~\cite{gu2021efficiently,smith2022simplified} has revitalized RNNs, addressing critical limitations that have hindered the effectiveness of traditional RNNs. These models have achieved state-of-the-art performance in the Long Range Arena benchmark~\cite{Tay_Dehghani_Abnar_Shen_Bahri_Pham_Rao_Yang_Ruder_Metzler_2020}. They have been applied to speech enhancement in \cite{ku2023multi} combined with the U-Net structure.

In contrast to CNNs, Spiking Neural Networks (SNNs) serve as the computational foundation underlying the functionality of neurobiological systems. They replicate neural activity patterns present in biological brains, utilizing spiking neurons that convey information through discrete pulses or spikes. The fusion of spike-based computing with neuromorphic hardware holds considerable promise for energy-efficient applications. Numerous studies have highlighted the efficacy of integrating SNNs with deep-learning methodologies~\cite{wu2018spatio, neftci2019surrogate}.
Efforts have been made to employ SNNs in the context of speech enhancement~\cite{riahi2023single, wall2016recurrent,xing2019noise, timcheck2023intel}. Yannan et al. propose a shallow lateral inhibitory SNN with spectrogram-based rate 
coding~\cite{xing2019noise}. Julie et al. enhance temporal correlation among similar frequency bands and eliminate irrelevant noise sources by adaptively configuring its connectivity for different acoustic environments~\cite{wall2016recurrent}. However, these two models do not incorporate learning mechanisms and their practical performance remains to be tested.

Spiking-UNet~\cite{riahi2023single} is a recently proposed model that integrates UNet with SNNs for single-channel speech enhancement. Despite its potential, the model's evaluation has been limited, and it continues to grapple with the computational demands associated with Unet.
Jonathan et al. employ a sigma-delta neural network (SDNN)~\cite{timcheck2023intel}, which is a modified version of the traditional feedforward ReLU neural network design. This SDNN capitalizes on sparse message transmission using graded spikes and stateful neurons. However, none of these endeavors have succeeded in striking a balance between performance and computational efficiency.

In this paper, we introduce a lightweight SNN-based model named "Spiking Structured State Spaces for Sequence Modeling"~(Spiking-S4), tailored for speech enhancement. Our contribution encompasses two pivotal aspects.

Firstly, we pioneer the combination of the structured state space model and spiking neural networks within the domain of speech enhancement.

Secondly, our "Spiking-S4" model demonstrates competitiveness with state-of-the-art techniques while significantly reducing computational overhead.

%% file: sections/pre.tex
\vspace{-2mm}
\section{Preliminaries}
\subsection{Spiking neural networks}

\begin{figure*}[bt]
\centering
\includegraphics[width=1.\textwidth]{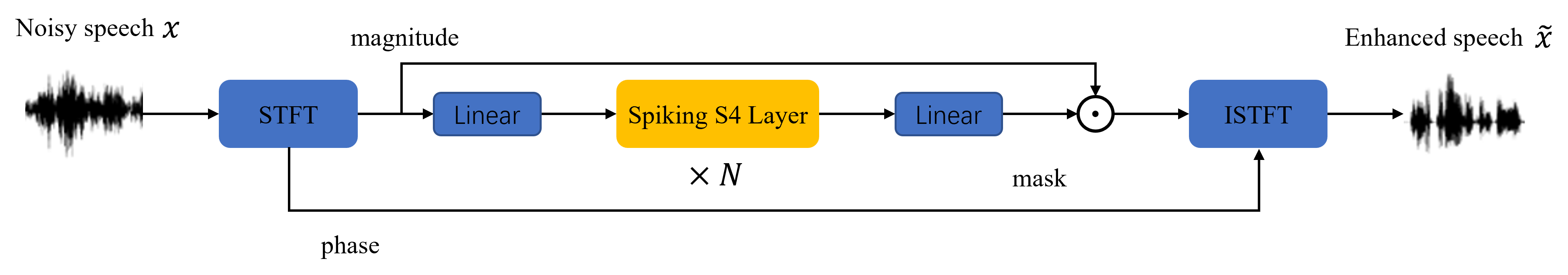} 
\caption{The overall framework}
\label{fig1}
\end{figure*}
We use the Leaky Integrate-and-Fire (LIF) model ~\cite{gerstner2014neuronal} as the spiking neuron in this work. It can be written in the discrete form of 
\begin{equation}
\label{eq:lif}
\begin{split}
&U(t) = U(t-1) +\alpha(O(t) - (V(t-1)-V_\textrm{reset}))\\
&S(t) = \Theta[t](U(t)-V_{\textrm{threshold}})\\
&V(t)=U(t)\cdot(1-S(t)) + V_{\textrm{reset}}\cdot S(t)
\end{split}
\end{equation}
where $\Theta$ denotes the Heaviside step function, $\alpha$ a decay factor, $V$ the membrane potential of the neuron, $S$ the spiking tensor, $O$ the output of the previous layer or initial input, $V_{\text{threshold}}$ the firing membrane threshold and $V_{\text{reset}}$ the reset potential.

Training SNNs with conventional gradient descent optimization methods is challenging due to the discrete nature of spiking signals, which hinders gradient propagation. To overcome this difficulty, researchers have proposed the Surrogate Gradient method~\cite{neftci2019surrogate}, which allows for effective backpropagation training in discrete spiking neural networks.

\subsection{Structured state space model}
Given an input scalar function $u(t)$, the continuous time-invariant SSM is defined by the following first-order differential equation that maps $u(t)$ to the output $y(t)$.
\begin{equation}
x^{\prime}(t) =\boldsymbol{A} x(t)+\boldsymbol{B} u(t), \quad y(t)  =\boldsymbol{C} x(t)+\boldsymbol{D} u(t)
\end{equation}

Extensive results show that initializing matrix $A$ with the HIPPO matrix~\cite{gu2020hippo} enables the SSM to effectively capture long-term dependencies. The $D$ can be considered as a parameter-dependent skip-connection. Therefore, we follow previous works~\cite{gu2021efficiently,gu2020hippo,smith2022simplified} and omit D from the SSM equation.

The SSM is then discretized using bilinear or zero-order hold (ZOH) methods, resulting in
\begin{equation}
\mathbf{x}_k=\bar{\mathbf{A}} \mathbf{x}_{k-1}+\overline{\mathbf{B}} \mathbf{u}_k, \quad \mathbf{y}_k=\overline{\mathbf{C}} \mathbf{x}_k
\end{equation}
and 
\begin{equation}
    \begin{aligned}
& \overline{\boldsymbol{A}}=(\boldsymbol{I}-\Delta / 2 \cdot \boldsymbol{A})^{-1}(\boldsymbol{I}+\Delta / 2 \cdot \boldsymbol{A}) \\
& \overline{\boldsymbol{B}}=(\boldsymbol{I}-\Delta / 2 \cdot \boldsymbol{A})^{-1} \Delta \boldsymbol{B} \\
&\overline{\boldsymbol{C}}=\boldsymbol{C}
\end{aligned}
\end{equation}
The S4 model incorporates a low-rank correction to regularize matrix A, facilitating stable diagonalization and transforming the SSM into a familiar convolutional computation with a Cauchy kernel.
\begin{equation}
y_t=\sum_{i=0}^t \overline{K_i} u_{t-i}, \quad \overline{K_i}=\bar{C} \bar{A}^{i-1} \bar{B}, \quad y=\overline{K} * u
\end{equation}

%% file: sections/method.tex
\vspace{-2mm}
\section{Method}
\subsection{Overview}
As shown in ~\cref{fig1}, the noisy speech signal is first transformed into the time-frequency domain using the Short-Time Fourier Transform (STFT) layer. Then, the magnitude is fed into the linear encoder to generate the input $u$ of shape ${K\times L}$ for the spiking S4 layers, where $K$ is the length of the input sequence and $L$ is the independent
S4 kernel number of each sequence element. Next it is passed to $N$ spiking S4 layers and a linear decoder, which produces a magnitude mask $\hat{M}$. This mask is then multiplied with the original magnitude to obtain the denoised magnitude. Finally, the denoised magnitude and the phase information are combined and converted back to the time domain using Inverse Short-Time Fourier Transform (ISTFT) layer. 

The STFT layer and ISTFT layer both involve precomputing Fourier coefficients and then freezing them as network weights, thereby excluding them from network training. This approach enables end-to-end computation within the network, greatly reducing the reliance on specific types and performance of computational resources. It also facilitates more efficient utilization of Graphics Processing Units~(GPUs) and other parallel computing resources, ultimately enhancing the overall model training and inference speed.

\subsection{Spiking S4 Layer}
\begin{figure*}[bt]
\centering
\includegraphics[width=1.\textwidth]{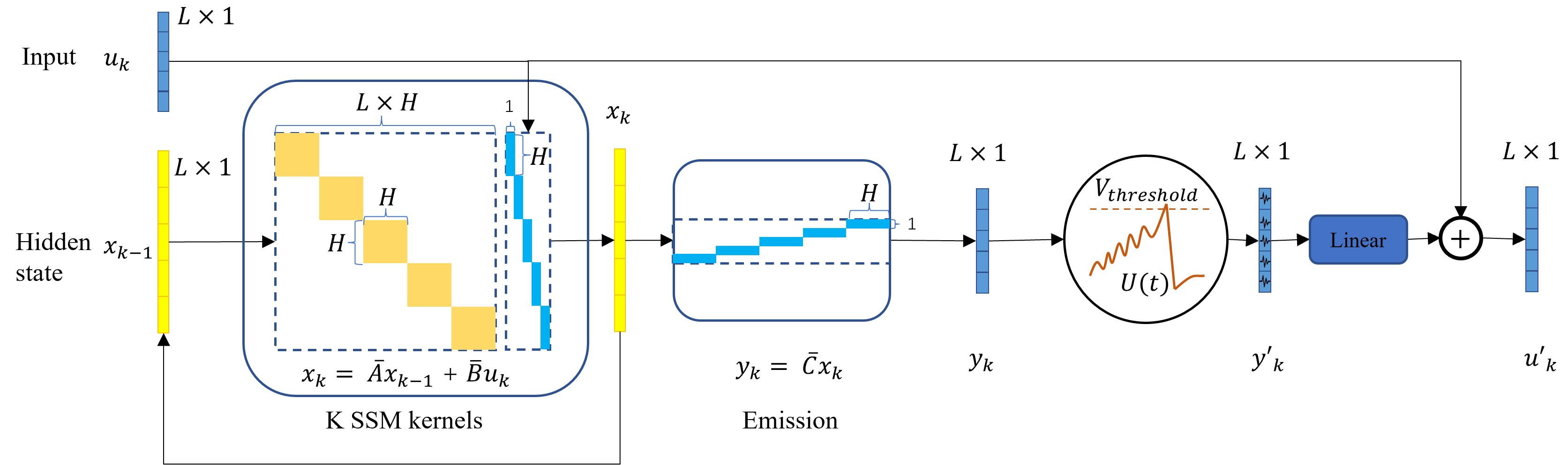} 
\caption{The spiking S4 layer}
\label{fig2}
\end{figure*}

Here, we show the recurrent mode of the Spiking S4 Layer, which means one step in, one step out. In practice, we employ the convolution mode which means all the timesteps are fed in concurrently and advanced techniques like parallel scan~\cite{gu2021efficiently} can be leveraged for acceleration.
As shown in \cref{fig2}, each step of the encoded feature is first passed to $L$ independent S4 kernels with the hidden size of $H$. Then it is passed to an emission layer,  and a LIF node which collects input signals gradually, accumulating them over a duration of time, and generates a spike when the membrane potential reaches a predefined threshold.
Finally, the spikes are fed into a linear decoder to be restored back to the real domain.

To mitigate information loss, a shortcut connection is incorporated.
We follow ~\cite{fang2021incorporating} and render both the synaptic weights and membrane time constants as learnable parameters.

\subsection{Loss function}
The loss function consists of two terms. The first term is the negative Scale-Invariant Signal-to-Noise Ratio (SI-SNR),
\begin{equation}
L_\text{sisnr}=-10 \log _{10} \frac{\left\|s_{\text {target }}\right\|^2}{\left\|e_{\text {noise }}\right\|^2},
\end{equation}
where $s_{\text {target }}:=\frac{\langle\hat{s}, s\rangle s}{\|s\|^2}$ and 
$e_{\text {noise }}:=\hat{s}-s_{\text {target }}$.

The second term is a mean square error (MSE) loss between the predicted magnitude mask $\hat{M}$ and the ground truth magnitude mask $M$,
\begin{equation}
L_{\text{mask}} = MSE(M, \hat{M})
\end{equation}

The overall loss function is,
\begin{equation}
\label{eq:loss}
L = L_\text{sisnr} + \lambda L_{\text{mask}}
\end{equation}

%% file: sections/exp.tex
\vspace{-2mm}
\section{Experiments}
\label{sec:exp}
\subsection{Dataset}
We use the Deep Noise Suppression (DNS) challenge dataset and a smaller dataset called Voice-Bank+Demand~\cite{valentini2017noisy} for evaluation

The DNS challenge~\cite{timcheck2023intel} aims to foster innovation in the realm of noise suppression, a critical component for achieving high-quality speech perception. To validate our approach, we employ the dataset for Intel DNS Challenge 2023 - Main (Real-Time) Track. It is in full-band format, comprising two subsets: clean full-band and noise full-band. It has a total size of 892 GB, of which 827 GB is allocated to clean full-band data and 58 GB to noisy full-band data. By using the official script, we synthesize a 500-hour dataset. Each dataset instance consists of 60000 samples, with each sample containing three audio files: clean audio file, noise audio file, and noisy audio file. The duration of each audio file is 30 seconds, with a sampling rate of 16 kHz. We allocate 80\% of the 60,000 samples to the training set and the remaining 20\% to the validation set.

Our evaluation metrics for testing include SI-SNR, Perceptual Evaluation of Speech Quality~(PESQ), Short-Time Objective Intelligibility~(STOI)~\cite{hu2007evaluation}, and Deep Noise Suppression Mean Opinion Score~(DNSMOS)~\cite{timcheck2023intel}.

The Voice-Bank+Demand dataset comprises 11,572 training pairs (from 28 speakers) and 824 testing pairs (from 2 speakers), serving as a relatively compact evaluation dataset. It has four evaluation metrics: Wide-Band PESQ (WB-PESQ), Composite Speech Intelligibility Grade (CSIG), Composite Background Noise Grade (CBAK), and Composite Overall Quality Grade (COVL)~\cite{hu2007evaluation}.

\subsection{Implementation details}
The STFT window length is set as 512. 
We train the models on a single A100 GPU and use the Rectified Adam~(RAdam) optimizer with a learning rate 0.001, and train the models for 50 epochs with a batch size of 128. 

The $V_\text{threshold}$, $V_\text{reset}$ and $\alpha$ in~\cref{eq:lif} is set as 1, 0, 2 respectively. We use the Atan function as the surrogate gradient function. The number of spiking S4 layers is set as 4 and the hidden size $H$ is 256. The $\lambda$ in~\cref{eq:loss} is 0.001.

\begin{table}[t]
\small
\caption{Results on DNS Challenge 2023 validation set and test set. The ANN and SNN-based models are separated by a horizontal line.}
\label{tab:dns}
  \centering
  \resizebox{0.75\linewidth}{!}{
  \begin{tabular}{c|c|c|c|c|c|c}
  \toprule
  Method&SISNR&PESQ&STOI&\multicolumn{3}{c}{DNSMOS}\\
  \midrule
   \multicolumn{7}{c}{Validation set}\\
   \midrule
  Wave-U-Net~\cite{macartney2018improved}&13.70&1.80&0.88&2.91&3.15&3.52\\

  S4~\cite{gu2021efficiently}&14.82&1.99&0.89&2.93&3.23&3.91\\
  FRCRN~\cite{zhao2022frcrn}&\textbf{15.51}&2.50&\textbf{0.92}&\textbf{3.09}&\textbf{3.41}&\textbf{3.96}\\
  \midrule
    Sigma-Delta~\cite{timcheck2023intel}&11.7&1.69&0.86&2.67&3.17&3.44\\
    Spiking-S4&14.42&\textbf{2.73}&0.89&2.85&3.21&3.74\\

  \midrule
  \multicolumn{7}{c}{Test set}\\
  \midrule
      Wave-U-Net~\cite{macartney2018improved}&13.90&1.85&1.02&3.01&3.25&3.65\\      S4~\cite{gu2021efficiently}&15.01&2.76&0.89&2.93&3.24&3.89\\
      FRCRN~\cite{zhao2022frcrn}&\textbf{15.67}&2.52&\textbf{0.92}&\textbf{3.08}&\textbf{3.41}&\textbf{3.95}\\
      \midrule
       Sigma-Delta~\cite{timcheck2023intel}&11.21&2.43&0.86&2.68&3.14&3.51\\
    Spiking-S4&14.58&\textbf{2.75}&0.89&2.85&3.21&3.74\\

  \bottomrule
\end{tabular}
}
\end{table}
\vspace{-2mm}
\subsection{Results}
\subsubsection{Results on DNS Challenge 2023 dataset}

We evaluate our spiking S4 model on the DNS challenge dataset 2023 validation set and test set. 
We compare spiking S4 and its ANN equivalent with the Intel DNS Challenge baseline neuromorphic model Sigma Delta Network~\cite{timcheck2023intel} and two open-sourced performant ANN models Wave-U-Net~\cite{macartney2018improved} and FRCRN~\cite{zhao2022frcrn}.
As shown in ~\cref{tab:dns}, S4 and Spiking S4 are competitive in ANN-based and SNN-based groups, respectively. FRCRN is based on the Complex-Unet and recurrent structure and achieves the best performance among the ANN models but incurs high training and inference costs. S4 is the closest to FRCRN with a much lower computation cost. For the SNN groups, our spiking S4 is slightly inferior to its ANN equivalent but clearly outperforms the Sigma-Delta network in all the indicators.

\subsubsection{Results on VoiceBank+Demand dataset}
VoiceBank+Demand is a relatively small dataset. As many works report results on this dataset, we can have more comparisons.
We show the results on the VoiceBank+Demand dataset in ~\cref{tab:voice bank}. As we can see, the S4 leads in CSIG and Spiking-S4 ranks the first in WB-PESQ and COVL.

\subsubsection{Computation cost}
To compare the computation cost, we list the model parameter numbers and FLOPs (the number of floating-point operations by forwarding a single sample in the network) in ~\cref{tab:param}.
Our Spiking-S4 model has the fewest parameters (0.53M) and FLOPs ($1.50\times10^9$) among all models, even fewer than the Intel DNS Challenge baseline solution Sigma-Delta Network~\cite{timcheck2023intel}.

\begin{table}[t]
\caption{Results on Voice-Bank+Demand dataset. The ANN and SNN-based models are separated by a horizontal line.}
\label{tab:voice bank}
  \centering
  \resizebox{0.7\linewidth}{!}{
  \begin{tabular}{c|c|c|c|c}
  \toprule
  Method&WB-PESQ&CSIG&CBAK&COVL\\
  \midrule
  Wave-U-Net~\cite{macartney2018improved}&3.25&4.20&3.61&3.30\\
  GaGNet~\cite{li2022glance}&2.94& 4.26 &3.45 &3.59\\
  MetricGAN+~\cite{fu2021metricgan+}&3.15& 4.14 &3.16& 3.64\\
  PERL-AE~\cite{kataria2021perceptual}&3.17& 4.43 &3.53& 3.83\\
  FRCRN~\cite{zhao2022frcrn}&3.21&4.23&3.64&3.37\\
  S4~\cite{gu2021efficiently}&3.38&\textbf{4.93}&2.63&4.30\\
  \midrule
   Sigma-Delta~\cite{timcheck2023intel}&3.20&4.89&2.59&4.15\\
  Spiking-S4&\textbf{3.39}&4.92&\textbf{2.64}&\textbf{4.31}\\

  \bottomrule
\end{tabular}
}
\end{table}

\begin{table}[h]
\caption{Comparision of parameters and FLOPs.}
\label{tab:param}
  \centering
  \resizebox{0.6\linewidth}{!}{
  \begin{tabular}{c|c|c}
  \toprule
  Method&Parameter&FLOPs\\
  Wave-U-Net~\cite{macartney2018improved}&70.1M&$3.36\times 10^{10}$\\
  GaGNet~\cite{li2022glance}&5.9M&$8.13\times 10^9$\\
  Sigma-Delta~\cite{timcheck2023intel}&0.53M&$1.97\times 10^9$\\
FRCRN~\cite{zhao2022frcrn}&14.0M&$1.13\times10^{12}$\\
  S4\cite{gu2021efficiently}&0.79M&$2.48\times 10^9$\\
  Spiking-S4&\textbf{0.53M}&\textbf{$1.50\times10^9$}\\
  \bottomrule
\end{tabular}
}
\end{table}

%% file: sections/con.tex
\vspace{-2mm}
\section{Conclusion}
\label{sec:con}
In conclusion, our paper introduces Spiking-S4, a lightweight SNN-based model designed for speech enhancement. It is the first work that combines the structured state space model  with the spiking neural networks and apply it to speech enhancement. Evaluation on two datasets demonstrates that our Spiking-S4 achieves competitive results with the ANN models while showing superior computation efficiency.